\documentclass[twocolumn,aps,prd,preprintnumbers,showpacs,superscriptaddress,nofootinbib,amsmath,amssymb,floats,floatfix,showkeys,notitlepage,longbibliography]{revtex4-1}
\usepackage{comment}
\usepackage{graphicx}
\usepackage{subfigure}
\usepackage{palatino}
\usepackage[commandnameprefix=always]{changes}
\usepackage{hyperref}
\hypersetup{colorlinks=true,linkcolor=blue,urlcolor=blue,citecolor=blue}
\usepackage[toc,page]{appendix}
\usepackage[normalem]{ulem}

\usepackage{orcidlink}
\usepackage{lipsum}
\usepackage{graphicx}
\usepackage{subfigure}
\usepackage{palatino}
\usepackage{sans}
\usepackage{adjustbox}
\usepackage{latexsym}
\usepackage{amsmath}
\usepackage{amssymb}
\usepackage{amsfonts}
\usepackage{dcolumn}
\usepackage{bm}
\usepackage{tikz}
\usepackage{bigints}
\usepackage{array,tabularx,multirow,booktabs}
\usepackage{longtable}
\usepackage[tracking=true]{microtype}
\SetTracking{}{500}
\SetTracking{encoding={*}, shape=sc}{40}
\UseRawInputEncoding
\allowdisplaybreaks
\usepackage{color}
\usepackage[table]{xcolor}

\begin{document} \sloppy

\title{Where Thermodynamics Meets Geometry: Critical-Radius Coincidences in Confining-NED Black Holes with Barrow Entropy}

\author{Erdem Sucu \orcidlink{0009-0000-3619-1492}}
\email{erdemsc07@gmail.com}
\affiliation{Eastern Mediterranean University, Physics Department, Famagusta, 99628 North Cyprus, via Mersin 10, T\"urkiye}

\author{\.{I}zzet Sakall{\i} \orcidlink{0000-0001-7827-9476}}
\email{izzet.sakalli@emu.edu.tr}
\affiliation{Eastern Mediterranean University, Physics Department, Famagusta, 99628 North Cyprus, via Mersin 10, T\"urkiye}

\begin{abstract}
We study a static, spherically symmetric black hole obtained from Einstein gravity coupled to a nonlinear electrodynamics model with a quark--antiquark confinement interaction. The metric extends Reissner--Nordstr\"om by a logarithmic correction controlled by $\zeta$, modifying both horizon structure and the near-singularity regime. The Hamilton--Jacobi tunneling method for Dirac fermions yields the Hawking temperature; the $\zeta$-dependent terms suppress the small-horizon divergence and signal a remnant. Quantum-gravitational fluctuations are incorporated through Barrow entropy with deformation index $\Delta$. Within the extended phase space we compute the internal energy, free energy, pressure, heat capacity, isothermal compressibility, and Joule--Thomson coefficient. The heat capacity locates $\Delta$-dependent stability regions; the compressibility stays negative across the domain analysed here, marking a mechanically rigid phase with no van der Waals criticality in this branch. The central result is a quadruple coincidence: the peak Hawking temperature, the heat-capacity divergence, the Joule--Thomson inversion, and the zero of the radial tidal force all sit at one radius $r_\star$ defined by $A''(r_\star)=0$, while the extremal horizon and the angular tidal-force zero coincide via $A'(r_h)=0$. These reduce the full critical-point analysis to two scalar equations on $A(r)$. Geometric tidal accelerations are mapped against the thermodynamic critical curves. Event Horizon Telescope observations of Sgr~A* translate into a constraint $\zeta\lesssim 0.7$ at $Q/M=0.5$, leaving a finite window open. The confinement term induces observable corrections to geodesic deviation.
\end{abstract}

\date{\today}


\maketitle

\section{Introduction}\label{sec1}

Black holes (BHs) remain the most extreme objects predicted by General Relativity (GR), with curvature so concentrated that classical descriptions break down at their centers~\cite{giddings1995black,gregory2025black}. Schwarzschild and Reissner--Nordstr\"om (RN) geometries, despite their predictive successes, retain a central singularity that signals the limits of classical physics and motivates either a modified gravitational sector or a modified matter content~\cite{melchor2024semiclassical,harada2025singularity}. The latter route, which we follow here, employs Nonlinear Electrodynamics (NED)~\cite{bronnikov2023regular,gaete2017note,mazharimousavi2023quark,azreg2025two}, the framework originally constructed to remove the divergent self-energy of point charges and now widely used as a mechanism for replacing the curvature singularity with a regular core~\cite{yang2022dyonically,bakopoulos2024exact,sucu2025charged}.

Within this framework, a confinement-type NED model inspired by the quark--antiquark interaction in quantum chromodynamics~\cite{mazharimousavi2023quark} produces an exact static spherically symmetric solution. The resulting Confined Nonlinear Electrodynamics (CNED) geometry deforms the RN background by a logarithmic correction controlled by the confinement parameter $\zeta$. Most prior studies of confinement-type NED black holes have focused on the deflection angle and lensing~\cite{sucu2024effect}, leaving the joint quantum and thermodynamic structure of the geometry comparatively unexplored. We address that gap here.

The semiclassical content is treated via the Hamilton--Jacobi tunneling method for Dirac fermions~\cite{gecim2015dirac,gecim2018quantum}. Tunneling formulations of Hawking emission, pioneered by Parikh and Wilczek~\cite{parikh2000hawking} and now developed in many directions~\cite{eslamzadeh2025brief,nozari2009parikh,tan2025black,sucu2025quantumCPC}, reduce horizon thermodynamics to a near-horizon WKB problem and connect the radiation temperature directly to the surface gravity. Recent applications of the same machinery to regular black holes, modified-gravity backgrounds, and Lorentz-violating sectors illustrate the breadth of the approach~\cite{albadawi2025frolov,ahmed2025adsbs,ahmed2025kalbramond}. The Hawking temperature $T_H$ as a function of the horizon radius $r_h$ and the confinement parameter $\zeta$ exhibits a finite maximum, with the small-horizon divergence suppressed by the logarithmic correction. This behavior is consistent with the existence of a quantum remnant.

Quantum-gravitational fluctuations of the horizon are next incorporated through Barrow entropy~\cite{capozziello2025barrow,barrow2021big,barrow2020area}, which models the event horizon as a fractal surface characterized by the deformation index $\Delta\in[0,1]$. Barrow entropy belongs to a wider family of non-extensive horizon entropies that includes the Tsallis--Cirto, R\'enyi, and Sharma--Mittal forms; all of these have been used to model how a quantum-gravity-induced horizon roughening shifts the cosmological and black-hole thermodynamic balance~\cite{rani2023impact}. Working in the extended phase space (with vanishing cosmological constant), we compute the internal energy, Helmholtz free energy, pressure, heat capacity, isothermal compressibility, and Joule--Thomson coefficient. The heat capacity locates the stability regions and second-order phase transitions, and the Joule--Thomson coefficient identifies the cooling/heating crossover with a $\Delta$-dependent inversion radius. The isothermal compressibility tracks the mechanical response of the system under pressure variation and gives an independent check on the role of the confinement sector in the stability picture.

A geometric counterpart of the thermodynamic analysis is provided by radial geodesics and by the tidal forces seen in a static orthonormal tetrad comoving with a freely falling observer~\cite{crispino2016tidal,arora2023exploring,cordeiro2025free,lima2020tidal,lima2022tidal,toshmatov2023tidal}. The CNED-induced corrections to the tidal accelerations carry direct imprints of the confinement sector on local geodesic deviation.

A central new result of this paper is a set of algebraic coincidences linking the thermodynamic and geometric critical radii. Four physically distinct phenomena (the peak of $T_H$, the divergence of the heat capacity, the Joule--Thomson inversion, and the zero-crossing of the radial tidal force) all collapse onto a single characteristic radius $r_\star$, defined by $A''(r_\star)=0$. Independently, the extremal horizon ($T_H=0$) and the zero of the angular tidal force both satisfy $A'(r_h)=0$. These identities are verified analytically and tabulated for representative parameter sets in Sec.~\ref{sec45}.

We further translate the metric correction into an observational signature. The Sgr~A* shadow size measured by the Event Horizon Telescope (EHT) collaboration, combined with the analytic photon-sphere expression of the CNED geometry, constrains $\zeta$ to a finite window. The resulting bound is summarized in Sec.~\ref{sec65}. Many studies have addressed shadow-based parameter constraints in modified geometries~\cite{konoplya2025primary,ali2025influence,walia2024exploring,pedrotti2024quasinormal,sarkar2025deflection}; the influence of CNED on the shadow has not been quantified before to our knowledge.

The remainder of the paper proceeds as follows. Section~\ref{sec2} sets up the CNED background. Section~\ref{sec3} carries out the Dirac tunneling calculation and reads off $T_H$. Barrow thermodynamics is treated in Sec.~\ref{sec4}. The coincidence-radii result and the associated tables are presented in Sec.~\ref{sec45}. Radial geodesics follow in Sec.~\ref{sec5} and tidal forces in Sec.~\ref{sec6}. Section~\ref{sec65} discusses the EHT shadow constraint on $\zeta$, and Sec.~\ref{sec7} closes the paper with a synthesis and outlook.

\section{Review of NED black hole incorporating quark--antiquark confinement} \label{sec2}

We examine Einstein gravity coupled to a nonlinear electromagnetic sector representing a confinement-type interaction. The dynamics are derived from the action~\cite{mazharimousavi2023quark,sucu2025dynamics}
{\small\begin{equation}
S=\int d^{4}x\,\sqrt{-g}\Bigg[\frac{\mathcal{R}}{16\pi G}+\mathcal{L}\Bigg],
\end{equation}}
where $\mathcal{R}$ denotes the Ricci scalar curvature and the gravitational constant is set to $G=1$. The nonlinear electromagnetic Lagrangian is taken in the form
{\small\begin{equation}
\mathcal{L}
=-\frac{16\Big(3\sqrt{-2\mathcal{F}}+
\zeta\big(\zeta+\sqrt{\zeta^{2}+4\sqrt{-2\mathcal{F}}}\big)\Big)
\sqrt{-2\mathcal{F}}}{3\left(\zeta+\sqrt{\zeta^{2}+4\sqrt{-2\mathcal{F}}}\right)^{4}}\,
\mathcal{F},
\end{equation}}
where the Maxwell scalar invariant is
\begin{equation}
\mathcal{F}=\frac14 F_{\mu\nu}F^{\mu\nu}
=-\frac12\left(\frac{q}{r^{2}}+\frac{f}{r}\right)^{2},
\qquad f=\zeta\sqrt{q},
\end{equation}
and $\zeta$ represents the confinement parameter.

Variation of the action yields the Einstein equations,
\begin{equation}
G_{\mu}^{\ \nu}=8\pi T_{\mu}^{\ \nu},
\end{equation}
where the stress--energy tensor of the nonlinear electrodynamic field reads
\begin{equation}
T_{\mu}^{\ \nu}=\frac{1}{4\pi}
\Big(\mathcal{L}\delta_{\mu}^{\nu}-\mathcal{L}_{\mathcal{F}}
F_{\mu\lambda}F^{\nu\lambda}\Big),
\qquad
\mathcal{L}_{\mathcal{F}}\equiv\frac{\partial\mathcal{L}}{\partial\mathcal{F}}.
\end{equation}

A static and spherically symmetric configuration is obtained from the line element
\begin{equation}
ds^{2}=-A(r)dt^{2}+A(r)^{-1}dr^{2}
+r^{2}d\theta^{2}+r^{2}\sin^{2}\theta\,d\phi^{2},\label{metric}
\end{equation}
with the lapse function~\cite{sucu2025dynamics}
\begin{equation}
A(r)=1-\frac{2M}{r}+\frac{Q^{2}}{r^{2}}
-\frac{4Q\sqrt{Q}\,\zeta}{3r}\ln r.\label{metricfunc}
\end{equation}

The parameter $Q$ denotes the electric charge and $M$ is the mass parameter. Depending on the relative magnitudes of $\zeta$, $M$ and $Q$, the solution describes a black hole or a naked singularity. Analytical results for the shadow radius of the Reissner--Nordstr\"om case as a function of charge appear in~\cite{Zakharov:2014lqa,Zakharov:2005ek}. Two limits are useful for orientation. At $\zeta\to 0$ the metric reduces to RN; at the additional limit $Q\to 0$ it becomes Schwarzschild. The logarithmic factor in Eq.~\eqref{metricfunc} is therefore the algebraic carrier of the confinement physics, and most quantities computed below will be RN-like with calculable $\zeta$-corrections.

\section{Quantum Tunnelling of Dirac Fermions}\label{sec3}

Due to the spherical symmetry of the spacetime, the tunnelling of outgoing particles takes place along radial trajectories, so the motion can be restricted to the equatorial plane without loss of generality. Imposing
\begin{equation}
\theta=\frac{\pi}{2},
\end{equation}
removes the polar angular dependence and yields the effective $(2+1)$-dimensional submanifold
\begin{equation}
ds^{2}=-A(r)\,dt^{2}+A(r)^{-1}\,dr^{2}+r^{2}\,d\phi^{2}.
\end{equation}
This is a slice, not a model. The reduction does not correspond to a lower-dimensional gravitational theory; it only selects the equatorial plane where the tunnelling occurs, while the background geometry remains fully $(3+1)$-dimensional. The Dirac equation can then be formulated on this effective two-dimensional spatial slice for the purpose of describing the Hawking emission.

The Dirac equation in the reduced geometry takes the form~\cite{sahan2025quantum}
\begin{equation}
i\,\sigma^{\mu}(x)\left(\partial_{\mu}-\Gamma_{\mu}(x)\right)\Psi(x)
= m_{0}\Psi(x),
\end{equation}
where the curved-space gamma matrices are written in terms of the triads $e^{\mu}{}_{(i)}$ as
{\small\begin{equation}
\sigma^{\mu}(x)=e^{\mu}{}_{(i)}\,\sigma^{i},\qquad
e^{\mu}{}_{(i)}=\mathrm{diag}\!\left(\sqrt{A(r)},\,\frac{1}{\sqrt{A(r)}},
\,\frac{1}{r}\right).
\end{equation}}
Here $\Psi$ is a two-component spinor and the flat-space matrices are chosen as $\sigma^{0}=\sigma^{3}$, $\sigma^{1}=i\sigma^{1}$, $\sigma^{2}=i\sigma^{2}$~\cite{gecim2017dirac}. The spin connection reads~\cite{gecim2015dirac}
\begin{align}
\Gamma_{\mu}&=\frac{1}{4}g^{\lambda\alpha}\left(e^{i}_{\nu,\mu} e_{\alpha i}-\Gamma^{\alpha}_{\nu\mu}\right)s_{\lambda\nu},\\
s_{\lambda\nu}&=\frac{1}{2}[\sigma_{\lambda},\sigma_{\nu}],\\
g_{\mu\nu}&=e^{(i)}_{\mu}e^{(j)}_{\nu}\eta_{(i)(j)}.
\end{align}

For the CNED black hole, the nonzero spin connections are
\begin{equation}
\Gamma_{0}=-\frac{1}{4}A'\sigma_{3}\sigma_{1},\quad \Gamma_{1}=0,\quad \Gamma_{2}=-\frac{1}{2}\sqrt{A}\,\sigma_{1}\sigma_{2}.
\end{equation}

Inserting these into the Dirac equation yields coupled equations for $\Psi_{1,2}$. With the WKB ansatz
\begin{equation}
\Psi=\begin{pmatrix}D\\B\end{pmatrix}\exp\left(\frac{i}{\hbar}S(t,r,\phi)\right),
\end{equation}
one obtains the Hamilton--Jacobi equation,
\begin{equation}
\frac{1}{A}(\partial_t S)^2 - A(\partial_r S)^2 - \frac{1}{r^2}(\partial_\phi S)^2 - m_0^2=0,
\end{equation}
which separates as $S=-Et+j\phi+K(r)+C$. The radial action is
\begin{equation}
K_{\pm}=\pm\int \frac{\sqrt{E^2-A(m_0^2+j^2/r^2)}}{A}\,dr.
\end{equation}
Near the horizon $r_h$, the linearization $A(r)\simeq A'(r_h)(r-r_h)$ gives
\begin{equation}
K_{\pm}=\pm \frac{i\pi E}{A'(r_h)},\quad \mathrm{Im}\,S=2\,\mathrm{Im}\,K_{+}.
\end{equation}
The tunnelling probability is therefore
\begin{equation}
\Gamma=\exp\left[-\frac{4\pi E}{ A'(r_h)}\right],
\end{equation}
yielding the Hawking temperature
{\small\begin{equation}
T_H=\frac{A'(r_h)}{4\pi}=\frac{M}{2 \pi  r_h^{2}}-\frac{Q^{2}}{2 \pi  r_h^{3}}+\frac{Q^{\frac{3}{2}} \zeta  \ln\! \left(r_h\right)}{3 \pi  r_h^{2}}-\frac{Q^{\frac{3}{2}} \zeta}{3 \pi  r_h^{2}},\label{2222}
\end{equation}}
in agreement with the surface gravity and Euclidean methods, with no factor-of-two ambiguity. The first two terms reproduce the standard RN temperature; the remaining $\zeta$-dependent pieces are the CNED corrections. We note that the condition $T_H = 0$ is equivalent, by the surface-gravity identification, to $A'(r_h)=0$, an algebraic relation that will reappear in Sec.~\ref{sec6} as the angular tidal force zero-crossing.

The validity of the near-horizon expansion deserves brief comment. The WKB ansatz is controlled by the small parameter $\hbar/(L\,E)$ with $L$ a length scale set by the geometry, and the linearization $A(r)\simeq A'(r_h)(r-r_h)$ becomes exact at the horizon and remains accurate within a thin shell of width $\delta r \sim T_H^{-1}\,|A''(r_h)/A'(r_h)|^{-1}$. Outside that shell, higher-order terms reshape the barrier but do not affect the imaginary part of the radial action, which is determined entirely by the residue at $r_h$. The temperature given by Eq.~\eqref{2222} is therefore a near-horizon invariant. A complementary check is the bosonic tunneling calculation~\cite{gecim2018gup,gecim2017gup,gecim2020quantum,tekincay2021zitterbewegung} for scalar and vector probes, which returns the same $T_H$ via the same surface-gravity identification; the fermion result presented here is consistent with that line of work~\cite{sucu2025nonlinear,sucu2025quantumozcan,sakalli2025zitterbewegung}.

\begin{figure}[t]
    \centering
    \includegraphics[width=0.46\textwidth]{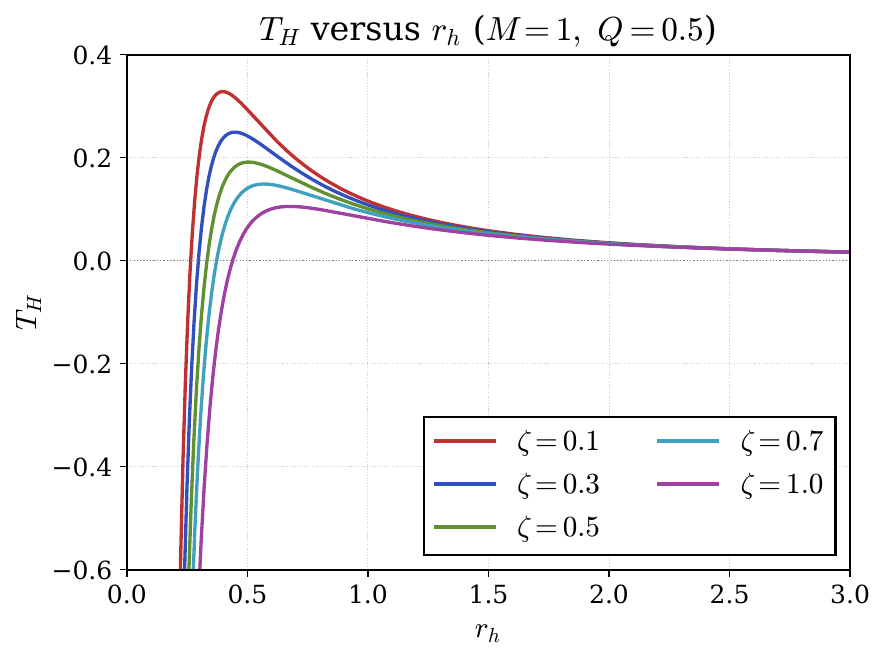}
    \caption{Hawking temperature $T_H$ versus horizon radius $r_h$ for various values of $\zeta$. The fixed parameters are $M = 1$ and $Q = 0.5$. The peak of each curve sits at the characteristic radius $r_\star$ analysed in Sec.~\ref{sec45}.}
    \label{tt}
\end{figure}

Figure~\ref{tt} shows $T_H$ as a function of $r_h$ for several values of $\zeta$. The structure is straightforward to read off. From Eq.~\eqref{2222}, the Schwarzschild contribution $M/(2\pi r_h^{2})$ and the repulsive RN piece $-Q^{2}/(2\pi r_h^{3})$ are supplemented by two $\zeta$-dependent logarithmic terms with opposite signs. As $r_h \to 0$ the negative logarithmic piece overcomes the others and forces $T_H$ to descend rather than diverge, signaling the possible presence of a remnant. At large $r_h$ all $\zeta$-corrections decay faster than the Schwarzschild term and the temperature curves approach the classical limit. Larger $\zeta$ shifts the temperature peak to larger $r_h$ and lowers its amplitude, so the confinement sector suppresses the rate of Hawking emission. The resulting non-monotonic profile, which rises from zero, peaks at a $\zeta$-dependent radius, and decreases at large $r_h$, signals an unstable--stable second-order phase transition. The peak position itself is the subject of Sec.~\ref{sec45}.

\section{Barrow entropy and extended-phase-space thermodynamics}\label{sec4}

Quantum gravitational fluctuations may render the horizon a fractal rather than a smooth two-sphere. Barrow parametrized this idea with a single index. The deformation is set by $0 \leq \Delta \leq 1$, with $\Delta=0$ recovering the standard Bekenstein--Hawking case and $\Delta=1$ corresponding to the maximally deformed configuration. In terms of the horizon radius and the geometric area $A=4\pi r^{2}$, the Barrow entropy reads~\cite{capozziello2025barrow}
\begin{equation}
S_{B}=(\pi r^{2})^{\,1+\frac{\Delta}{2}} .
\label{SB1}
\end{equation}
The parameter $\Delta$ encodes the cumulative effect of horizon-surface fluctuations without an extra dimensional constant. Equivalently, Eq.~\eqref{SB1} is the Tsallis--Cirto form with non-extensivity index $\delta_{T}=1+\Delta/2$, so the analysis below also covers the corresponding Tsallis--Cirto correction. The closely related R\'enyi and Sharma--Mittal entropies, briefly compared in Sec.~\ref{sec45}, give the same critical-radius structure but with different prefactors in the extensive sector.

To incorporate the geometric deformation, the differential first law $dE_{B}=T_{H}\,dS_{B}$ is applied with Eq.~\eqref{2222} substituted for $T_{H}$ and $S_{B}$ written as a function of $r_{h}$. The integration is nontrivial because the logarithmic confinement terms and the fractal factor $(r_{h}^{2})^{\Delta/2}$ contribute at the same order. The internal energy
\begin{equation}
E_{B}= \int T_H\, dS_B,
\label{eq:first_law_integration}
\end{equation}
where $T_H$ is given by Eq.~\eqref{2222}, then follows by substitution of the corrected entropy~\eqref{SB1}:
{\small
\begin{multline}
    E_B = \frac{\left(r_h^{2}\right)^{\frac{\Delta}{2}} \pi^{\frac{\Delta}{2}} Q^{\frac{3}{2}} \ln\! \left(r_h\right) \zeta}{3 \left(\Delta -1\right)}-\frac{2 \left(r_h^{2}\right)^{\frac{\Delta}{2}} \pi^{\frac{\Delta}{2}} Q^{\frac{3}{2}} \zeta}{3 \left(\Delta -1\right)}\\-\frac{2 \left(r_h^{2}\right)^{\frac{\Delta}{2}} \pi^{\frac{\Delta}{2}} Q^{\frac{3}{2}} \ln\! \left(r_h\right) \zeta}{3 \Delta  \left(\Delta -1\right)}+\frac{2 \left(r_h^{2}\right)^{\frac{\Delta}{2}} \pi^{\frac{\Delta}{2}} Q^{\frac{3}{2}} \zeta}{3 \Delta^{2} \left(\Delta -1\right)}\\+\frac{\left(r_h^{2}\right)^{\frac{\Delta}{2}} \pi^{\frac{\Delta}{2}} M}{2 \left(\Delta -1\right)}-\frac{\pi^{\frac{\Delta}{2}} Q^{2} \left(r_h^{2}\right)^{\frac{\Delta}{2}}}{r_h \left(\Delta -1\right)}-\frac{\left(r_h^{2}\right)^{\frac{\Delta}{2}} \pi^{\frac{\Delta}{2}} M}{\Delta  \left(\Delta -1\right)}\\+\frac{\left(r_h^{2}\right)^{\frac{\Delta}{2}} \pi^{\frac{\Delta}{2}} \Delta  Q^{\frac{3}{2}} \zeta  \ln\! \left(r_h\right)}{3 \left(\Delta -1\right)}-\frac{\left(r_h^{2}\right)^{\frac{\Delta}{2}} \pi^{\frac{\Delta}{2}} \Delta  Q^{\frac{3}{2}} \zeta}{3 \left(\Delta -1\right)}\\+\frac{\left(r_h^{2}\right)^{\frac{\Delta}{2}} \pi^{\frac{\Delta}{2}} Q^{\frac{3}{2}} \zeta}{3 \Delta  \left(\Delta -1\right)}+\frac{\left(r_h^{2}\right)^{\frac{\Delta}{2}} \pi^{\frac{\Delta}{2}} \Delta  M}{2 \left(\Delta -1\right)}-\frac{\Delta  \,\pi^{\frac{\Delta}{2}} Q^{2} \left(r_h^{2}\right)^{\frac{\Delta}{2}}}{2 r_h \left(\Delta -1\right)}
\end{multline}}

\begin{figure}
    \centering
    \includegraphics[width=0.46\textwidth]{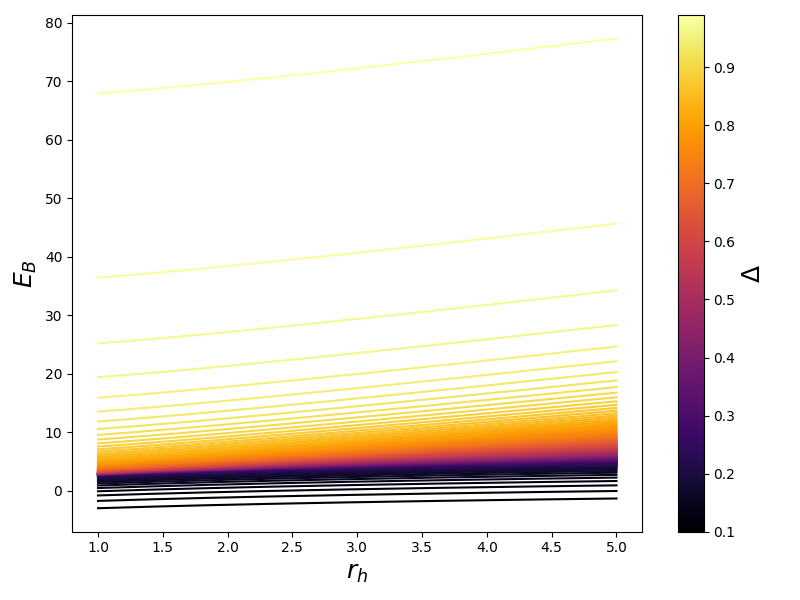}
    \caption{Internal energy $E_B$ as a function of $r_h$ for different Barrow indices $\Delta$, keeping $M = 1$, $Q = 0.5$, and $\zeta = 0.5$ fixed.}
    \label{ee}
\end{figure}

Figure~\ref{ee} shows the variation of $E_{B}$ with $r_{h}$ for several values of $\Delta$. The behavior is clean. Equation~\eqref{eq:first_law_integration} differs from the semiclassical expression through multiplicative factors of $(r_{h}^{2})^{\Delta/2}$, the $\ln r_{h}$ contributions, and the nonlinear $Q^{3/2}\zeta$ combinations of the CNED sector. Larger $\Delta$ produces a faster rise of $E_{B}$ with $r_{h}$, indicating that stronger fractal deformation increases the effective energy storage capacity. At small $r_{h}$ the curves separate more strongly, evidence that quantum corrections dominate near the microscopic regime; at larger $r_{h}$ they converge towards the standard Bekenstein--Hawking baseline. The growth is monotonic, so no thermodynamic instability appears in this sector; the slope shift captures the degree of departure from the classical case.

The Helmholtz free energy is defined by~\cite{sucu2025astrophysical}
{\small\begin{equation}
F_{B}=-\int S_B\, dT_H.
\label{eq:helmholtz_definition}
\end{equation}}
Evaluating this integral gives
{\small
\begin{multline}
    F_B= \frac{2 \left(r_h^{2}\right)^{\frac{\Delta}{2}} \pi^{\frac{\Delta}{2}} Q^{\frac{3}{2}} \ln\! \left(r_h\right) \zeta}{3 \left(\Delta -1\right)}-\frac{\left(r_h^{2}\right)^{\frac{\Delta}{2}} \pi^{\frac{\Delta}{2}} Q^{\frac{3}{2}} \zeta}{\Delta -1}\\+\frac{\left(r_h^{2}\right)^{\frac{\Delta}{2}} \pi^{\frac{\Delta}{2}} Q^{\frac{3}{2}} \zeta}{3 \Delta  \left(\Delta -1\right)}-\frac{2 \left(r_h^{2}\right)^{\frac{\Delta}{2}} \pi^{\frac{\Delta}{2}} Q^{\frac{3}{2}} \ln\! \left(r_h\right) \zeta}{3 \Delta  \left(\Delta -1\right)}\\+\frac{2 \left(r_h^{2}\right)^{\frac{\Delta}{2}} \pi^{\frac{\Delta}{2}} Q^{\frac{3}{2}} \zeta}{3 \Delta^{2} \left(\Delta -1\right)}+\frac{\left(r_h^{2}\right)^{\frac{\Delta}{2}} \pi^{\frac{\Delta}{2}} M}{\Delta -1}\\-\frac{3 \pi^{\frac{\Delta}{2}} Q^{2} \left(r_h^{2}\right)^{\frac{\Delta}{2}}}{2 r_h \left(\Delta -1\right)}-\frac{\left(r_h^{2}\right)^{\frac{\Delta}{2}} \pi^{\frac{\Delta}{2}} M}{\Delta  \left(\Delta -1\right)}
\end{multline}}

\begin{figure}
    \centering
    \includegraphics[width=0.46\textwidth]{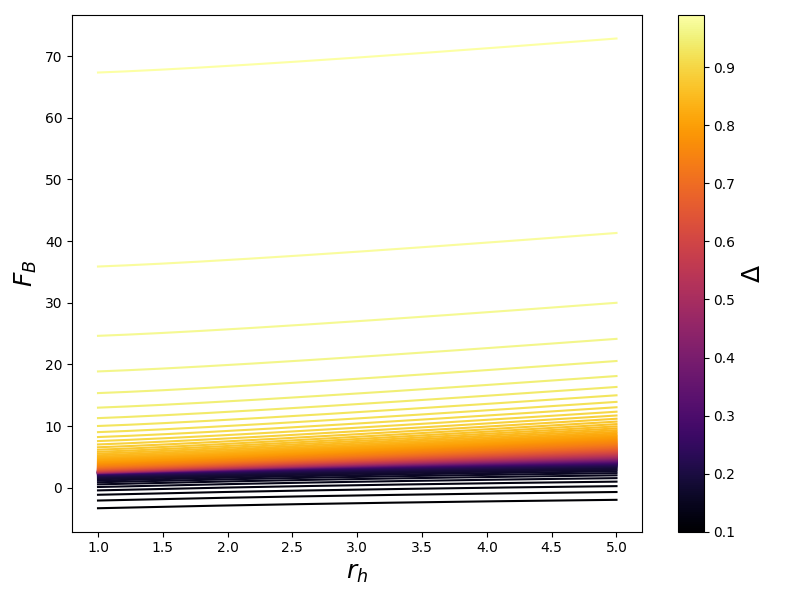}
    \caption{Behaviour of the Helmholtz free energy $F_B$ with respect to $r_h$ for several values of $\Delta$. The parameters are fixed to $M = 1$, $Q = 0.5$, and $\zeta = 0.5$.}
    \label{ff}
\end{figure}

Figure~\ref{ff} displays $F_{B}$ as a function of $r_{h}$ for different $\Delta$. Two ingredients matter. Equation~\eqref{eq:helmholtz_definition} depends on the fractal modification of the horizon through $(r_{h}^{2})^{\Delta/2}$ and on the nonlinear $\zeta Q^{3/2}$ terms inherited from the confinement sector. The small-$r_{h}$ behavior is sensitive to $\Delta$, where the free energy is dominated by the quantum fluctuations; as $r_{h}$ grows, the curves converge to the classical baseline. Larger $\Delta$ raises $F_{B}$, so a more deformed horizon requires a higher energetic cost to maintain thermodynamic equilibrium. The free energy is monotonic in $r_{h}$ and has no turning point, which indicates the absence of a first-order phase transition in this sector.

Within the extended thermodynamic picture, the pressure is
\begin{equation}
P_{B}= -\frac{dF_{B}}{dV},
\label{eq:pressure_definition}
\end{equation}
so that
{\small
\begin{multline}
    P_B=-\frac{3 \Delta  \,\pi^{\frac{\Delta}{2}} Q^{2} \left(r_h^{2}\right)^{\frac{\Delta}{2}}}{2 r_h^{2} \left(\Delta -1\right)}-\frac{\pi^{\frac{\Delta}{2}} \left(r_h^{2}\right)^{\frac{\Delta}{2}} Q^{\frac{3}{2}} \zeta}{r_h}\\+\frac{\pi^{\frac{\Delta}{2}} \left(r_h^{2}\right)^{\frac{\Delta}{2}} M}{r_h}+\frac{3 \pi^{\frac{\Delta}{2}} Q^{2} \left(r_h^{2}\right)^{\frac{\Delta}{2}}}{2 r_h^{2} \left(\Delta -1\right)}+\frac{2 \pi^{\frac{\Delta}{2}} Q^{\frac{3}{2}} \zeta  \ln\! \left(r_h\right) \left(r_h^{2}\right)^{\frac{\Delta}{2}}}{3 r_h}
\end{multline}}

\begin{figure}
    \centering
    \includegraphics[width=0.46\textwidth]{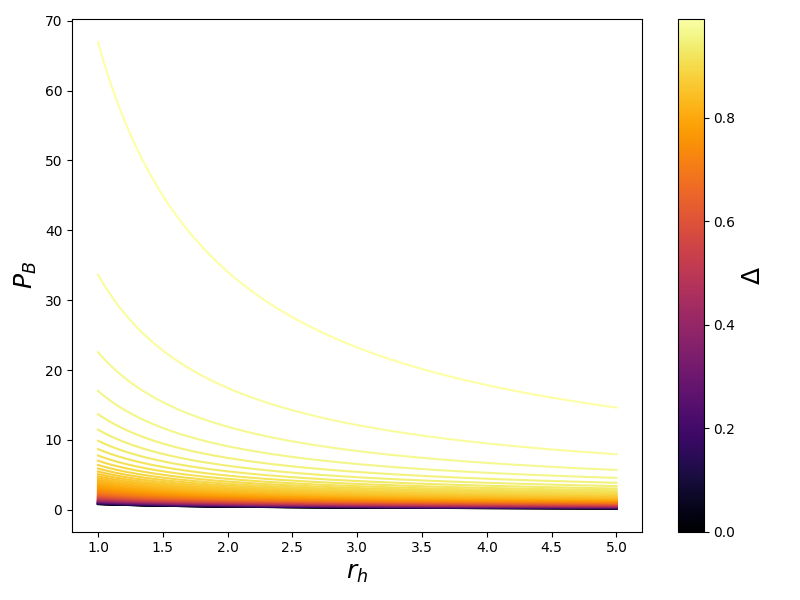}
    \caption{Pressure $P_B$ versus $r_h$ under variation of $\Delta$, evaluated for $M = 1$, $Q = 0.5$, and $\zeta = 0.5$.}
    \label{pp}
\end{figure}

Figure~\ref{pp} shows $P_{B}$ as a function of $r_{h}$ for several $\Delta$. The plot is layered. From Eq.~\eqref{eq:pressure_definition} the competing pieces are the $Q^{3/2}\zeta$ confinement terms, the $\ln(r_{h})$ corrections, and the Barrow factor $(r_{h}^{2})^{\Delta/2}$. The small-$r_{h}$ pressure is highly sensitive to $\Delta$: enhancement or suppression occurs depending on the strength of the quantum fluctuations. At larger radii the $\Delta$-dependence weakens and the classical behavior is recovered. Sign changes or extrema in $P_{B}$ at certain $\Delta$ values are indicators of thermodynamic instability or phase transitions. Larger $\Delta$ shifts these critical points and increases the effective vacuum repulsion, so the fractal deformation of the horizon affects the compressibility of the black hole~\cite{capozziello2025barrow}.

The heat capacity at constant volume locates the local stability of the system through
{\small\begin{equation}
C_{B} = T_H
\left(\frac{\partial S_B}{\partial T_H}\right).
\label{eq:heat_capacity_definition}
\end{equation}}
Using Eqs.~\eqref{SB1} and \eqref{2222}, one finds
{\small
\begin{equation}
\mathcal{D}(r_h;\zeta) 
=
4r_h\zeta
\left(\ln r_h-\frac{3}{2}\right)Q^{\frac{3}{2}}
+6Mr_h
-9Q^{2}.
\label{eq:DB}
\end{equation}

\begin{equation}
\begin{aligned}
C_B
&=
-\frac{
(2+\Delta)\,\pi^{1+\frac{\Delta}{2}}\,r_h^{2+\Delta}
}{
\mathcal{D}(r_h;\zeta) 
}
\\
&\quad \times
\left[
2r_h\zeta(\ln r_h-1)Q^{\frac{3}{2}}
+3Mr_h
-3Q^{2}
\right].
\end{aligned}
\label{eq:CB}
\end{equation}

\begin{figure}
    \centering
    \includegraphics[width=0.46\textwidth]{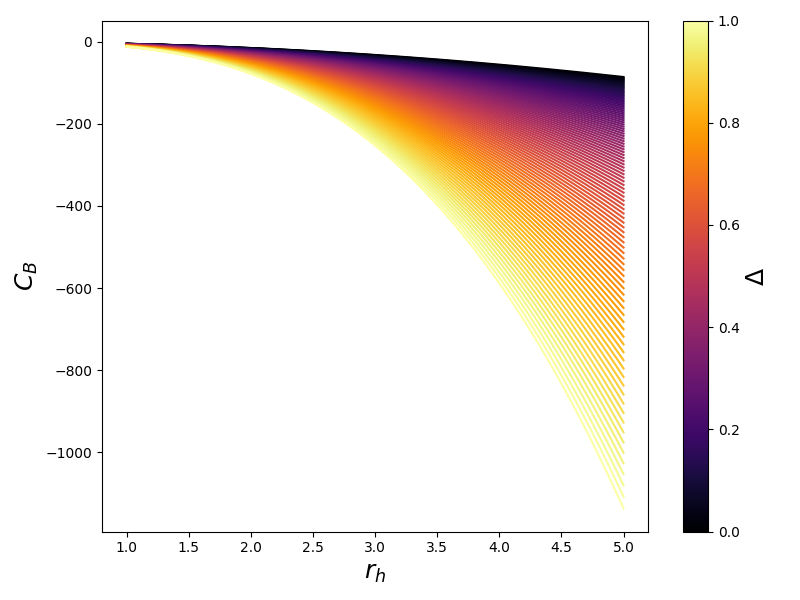}
    \caption{Heat capacity $C_B$ as a function of $r_h$ for different values of $\Delta$. Calculations are performed with $M = 1$, $Q = 0.5$, and $\zeta = 0.5$.}
    \label{cc}
\end{figure}

Figure~\ref{cc} shows $C_{B}$ as a function of $r_{h}$ for several $\Delta$. Positive values mark thermodynamically stable phases, while negative regions signal instability. The denominator of Eq.~\eqref{eq:CB}, namely $\mathcal{D}(r_h;\zeta) $ controls the divergences (where $C_{B}\to\pm\infty$) and identifies the second-order phase transitions. The Barrow index $\Delta$ enters only as an overall multiplicative factor $r_h^{2+\Delta}\pi^{1+\Delta/2}(2+\Delta)$, so the locations of the divergences are $\Delta$-independent and depend solely on $\zeta$. The stability window widens as $\Delta$ grows, while the unstable region migrates to smaller $r_{h}$, indicating that stronger fractal deformation enhances the thermal response and delays the onset of instability~\cite{rani2023impact,tekincay2021exotic}.

The Joule--Thomson coefficient, which measures the temperature change during isenthalpic expansion, is defined as~\cite{aydiner2025regular}
{\small\begin{equation}
\mu_{J}
    = \left( \frac{\partial T_H}{\partial P} \right)_{H}
    =  \displaystyle \frac{\partial T_H}{\partial r_{h}} 
           { \displaystyle \frac{\partial r_{h}}{\partial P_B} },
\label{eq:JT_definition}
\end{equation}}
so that
\begin{equation}
    \mu_{J}=\frac{\mathcal{A}}{\mathcal{B}}
\end{equation}
with
{\small
\begin{equation}
\begin{aligned}
\mathcal{A}
&=
4r_h
\left(r_h^{2}\right)^{-\frac{\Delta}{2}}
\pi^{-\frac{\Delta}{2}}
\\
&\quad \times
\left[
4r_h\zeta
\left(\ln r_h-\frac{3}{2}\right)Q^{\frac{3}{2}}
+6Mr_h
-9Q^{2}
\right].
\end{aligned}
\label{eq:A}
\end{equation}
and
{\small\begin{multline}
    \mathcal{B}= 4 r_h \left(\left(\Delta -3\right) \ln\! \left(r_h\right)-\frac{3 \Delta}{2}+\frac{11}{2}\right) \zeta  Q^{\frac{3}{2}}\\+6 M \left(\Delta -3\right) r_h-9 Q^{2} \left(\Delta -4\right)
\end{multline}}

\begin{figure}
    \centering
    \includegraphics[width=0.46\textwidth]{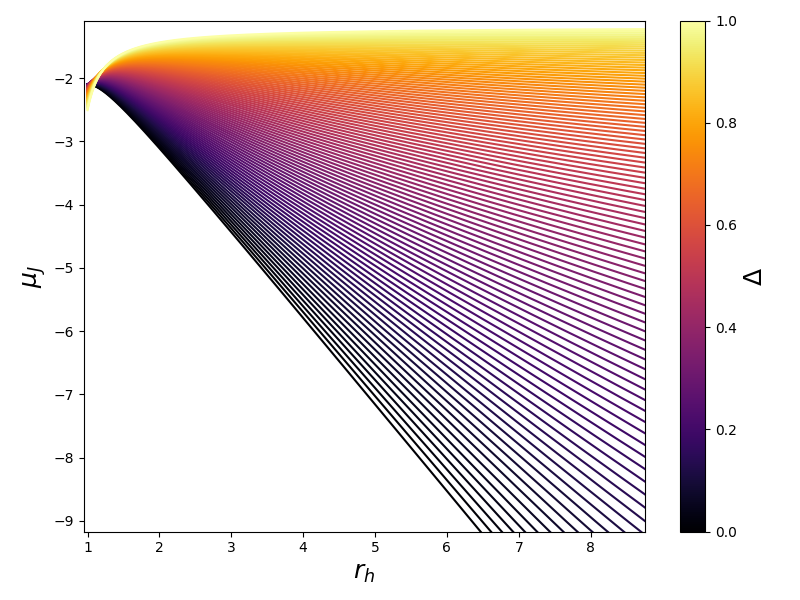}
    \caption{Joule--Thomson coefficient $\mu_J$ plotted against $r_h$ for distinct Barrow parameters $\Delta$ with $M = 1$, $Q = 0.5$, and $\zeta = 0.5$.}
    \label{jte}
\end{figure}

Figure~\ref{jte} shows $\mu_{J}$ as a function of $r_{h}$ for several $\Delta$. The structure is rich. The coefficient is sensitive to the interplay between the confinement-induced $Q^{3/2}\zeta$ sector and the fractal factor $(r_{h}^{2})^{\Delta/2}$, leading to inversion points where $\mu_{J}=0$. A positive $\mu_{J}$ marks cooling on expansion, a negative value marks heating, and the sign changes locate the inversion radius $r_{\rm inv}$. Comparing the numerator $\mathcal{A}$ with the denominator $\mathcal{D}$ in Eq.~\eqref{eq:DB} reveals that $\mathcal{A}\propto \mathcal{D}$ up to a $\Delta$-dependent prefactor that vanishes nowhere on the physical interval. Hence $\mu_{J}=0$ precisely at the heat-capacity divergence. Larger $\Delta$ shifts the inversion through the denominator $\mathcal{B}$, so quantum-gravitational fluctuations modulate, but do not displace, the inversion radius, which retains its $\Delta$-independent location. The behavior parallels van der Waals-like fluids and points to distinct thermodynamic regimes set by horizon geometry~\cite{li2018holographic,sekhmani2025thermodynamics}.

\paragraph*{Van der Waals-like phase structure.}
The pair $(T_H,P_B)$ at fixed $\Delta$ traces out an isotherm in the $P_B$--$r_h$ plane that has the qualitative shape of a non-ideal fluid. A turning point in $P_B$ along an isotherm flags a first-order transition; the inflection in $T_H$ along an isobar flags a second-order one. For Eqs.~\eqref{2222} and \eqref{eq:pressure_definition} the second-order critical point sits at $r_\star$ and shifts with $\zeta$ only, an observation that we elevate to a general statement in Sec.~\ref{sec45}. A Maxwell construction equalizing the chemical potential along the unstable branch is therefore feasible whenever a first-order coexistence appears at large $\Delta$; the qualitative picture is the standard small-/large-black-hole coexistence familiar from RN-AdS thermodynamics~\cite{li2018holographic}. We have not attempted a quantitative Maxwell construction here because the $\Lambda=0$ branch we analyse is free of first-order transitions, but the algebraic preparation is in place if a cosmological term is added in follow-up work.

\paragraph*{Isothermal compressibility.}
The heat capacity and the Joule--Thomson coefficient probe the thermal response of the geometry. Mechanical response is read off a third quantity, the isothermal compressibility $\kappa$. For an ordinary fluid, the sign and divergence structure of $\kappa$ mark the onset of mechanical instability and the location of critical points; the same statement carries over to the black-hole sector after the extended-phase-space identifications are made~\cite{mangut2025lorentz}. Within the Barrow framework we define
\begin{equation}
\kappa
=
-\frac{1}{V}
\left(
\frac{\partial V}{\partial P_B}
\right)_{T_H},
\label{eq:kappa_def}
\end{equation}
where $V$ is the thermodynamic volume conjugate to $P_B$. Because both $P_B$ and $V$ are functions of $r_h$ at fixed $\Delta$, the partial derivative is computed by the chain rule once the Barrow-corrected potentials of the previous subsections are substituted. Writing
\begin{equation}
\mathcal{K}_{\Delta}(r_h)
\,\equiv\,
(\Delta-3)\,\ln(r_h)
-\frac{3\Delta}{2}
+\frac{11}{2}
\label{eq:Bdelta}
\end{equation}
for the logarithmic combination that appears in the result, the compressibility reads
\begin{equation}
\kappa \,=\,
\frac{
72\,\pi^{1-\frac{\Delta}{2}}\, r_h^{4}\,
\left(r_h^{2}\right)^{-\frac{\Delta}{2}}
}{
4\,\zeta\, r_h\, \mathcal{K}_{\Delta}(r_h)\, Q^{3/2}
\,+\,
6 M\,(\Delta-3)\, r_h
\,-\,
9 Q^{2}\,(\Delta-4)
}.
\label{eq:kappa_final}
\end{equation}
The numerator is positive and carries the Barrow signature through the factor $(r_h^{2})^{-\Delta/2}$. The denominator mixes the nonlinear confinement piece $\zeta Q^{3/2}$ with the logarithmic correction inherited from the metric and with two terms linear in $\Delta$. In the $\Delta\to 0$ limit, the standard semiclassical compressibility is recovered; nonzero $\Delta$ amplifies the role of microscopic horizon fluctuations.

\begin{figure}
    \centering
    \includegraphics[width=0.46\textwidth]{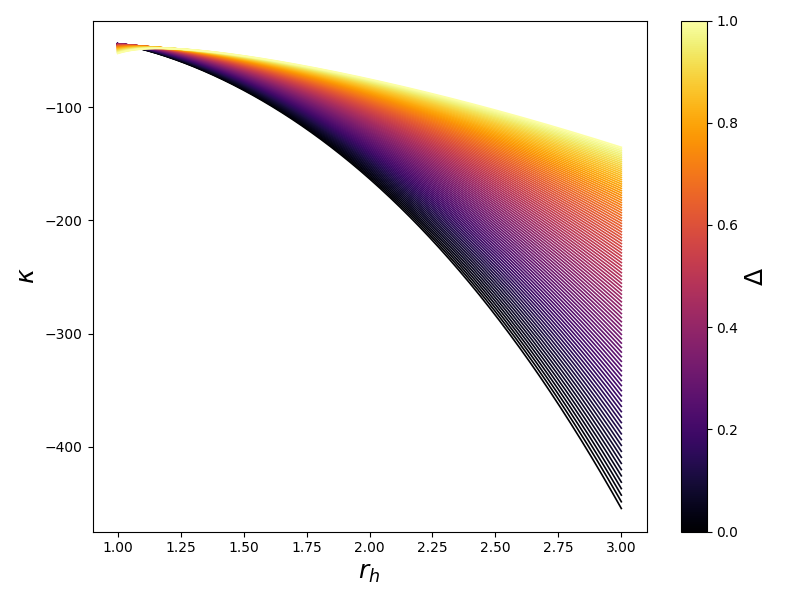}
    \caption{Isothermal compressibility $\kappa$ versus horizon radius for several values of the Barrow parameter $\Delta$, with $M=1$, $Q=0.5$, and $\zeta=0.5$.}
    \label{kappaaa}
\end{figure}

Figure~\ref{kappaaa} shows the behavior of $\kappa$ as a function of $r_h$ for several values of $\Delta$. The profile is smooth and monotonic across the displayed interval, with no divergence and no sign change. The denominator of Eq.~\eqref{eq:kappa_final} does not vanish on the plotted range, so the thermodynamic response stays regular and the system does not cross a critical point in this parameter window.

A common feature of all curves is the negative value of $\kappa$. In the black-hole reading, negative isothermal compressibility means that the effective thermodynamic volume contracts under an applied pressure increase, the response of a mechanically rigid configuration. The behavior is opposite to that of an ordinary extensive fluid, where positive $\kappa$ marks a stable equilibrium, and instead resembles the exotic response familiar from self-gravitating systems.

The Barrow parameter shifts the curves in a clean way. At fixed $r_h$, increasing $\Delta$ raises $\kappa$ toward zero, so the magnitude of the negative response shrinks. Fractal horizon deformation therefore softens the rigidity of the geometry. The shift is modest at small horizon radii. It grows as $r_h$ increases.

Below $r_h\sim 0.5$ all curves sit close together, and the quantum-geometric correction is mild. Above that radius the $\Delta$-branches fan out. The widening signals that the Barrow correction enters through higher-order scaling terms that gain weight at large $r_h$. No turning point appears. The monotone decrease of $\kappa$ with $r_h$ rules out van der Waals-type criticality in this branch.

The $\Delta\to 0$ limit recovers the standard Bekenstein--Hawking baseline, with the steepest decrease and the most rigid mechanical response. Larger $\Delta$ moderates the curve. Horizon fractality therefore lowers the sensitivity of the configuration to pressure variation. The Barrow deformation acts as a regulator of the response rather than as a generator of new critical structure; the confinement-induced logarithmic sector sets the global trend while the Barrow index sets the spread.

\section{Coincidence of thermodynamic and geometric critical radii}\label{sec45}

The above analysis exposes a pattern. The isothermal compressibility does not add a new critical radius in the parameter window of Fig.~\ref{kappaaa}, but it reinforces the stability picture obtained from the heat-capacity and Joule--Thomson sectors. The peak of the Hawking temperature, the divergence of the heat capacity, and the Joule--Thomson inversion are all controlled by the same algebraic object. We now make this explicit and add the geometric counterpart from tidal forces.

Differentiating Eq.~\eqref{2222} gives
\begin{equation}
\frac{\partial T_H}{\partial r_h}=\frac{A''(r_h)}{4\pi},
\label{eq:dTHdr}
\end{equation}
so the temperature peak sits at $A''(r_\star)=0$. A short algebraic step shows that the heat-capacity denominator~\eqref{eq:DB} satisfies
\begin{equation}
\mathcal{D}(r_h;\zeta)\;=\;-\frac{3}{2}\,r_h^{\,4}\,A''(r_h),
\label{eq:Dident}
\end{equation}
so $C_B\to\pm\infty$ at $A''(r_h)=0$ as well. Because $\mathcal{A}\propto\mathcal{D}$, the Joule--Thomson numerator also vanishes there, $\mu_J(r_\star)=0$. Finally, the radial tidal force on a freely falling observer is $-A''(r)/2$ in the orthonormal frame, so its zero-crossing coincides with $A''(r_h)=0$ as well. Four distinct physical phenomena therefore collapse onto a single characteristic radius:
{\small
\begin{equation}
\begin{aligned}
r_\star:\quad
\left.\frac{\partial T_H}{\partial r_h}\right|_{r_\star}=0
&\;\Longleftrightarrow\;
C_B\to\pm\infty
\\
&\;\Longleftrightarrow\;
\mu_J=0
\\
&\;\Longleftrightarrow\;
\left.
\frac{d^{2}\eta^{\hat r}}{d\tau^{2}}
\right|_{r_\star}=0 .
\end{aligned}
\label{eq:quad}
\end{equation}
Here, $r_\star$ simultaneously denotes the location of the peak Hawking
temperature, the second-order phase-transition point, the Joule--Thomson
inversion point, and the zero of the radial tidal-force component.

A parallel identity holds at one step lower in derivatives. The Hawking temperature itself vanishes at $A'(r_h)=0$, and the angular tidal force $-A'(r)/(2r)$ also vanishes there. So the extremal horizon coincides with the angular tidal force zero-crossing,
\begin{equation}
r_{\rm ext}:\quad T_H=0 \;\Longleftrightarrow\; A'(r_{\rm ext})=0 \;\Longleftrightarrow\;
\left.\frac{d^{2}\eta^{\hat\perp}}{d\tau^{2}}\right|_{r_{\rm ext}}=0.
\label{eq:pair}
\end{equation}

These identities are not coincidences in any physical sense; they follow from the surface-gravity definition of $T_H$ and the orthonormal-frame expression for tidal forces. What is nontrivial is that they reduce the eight separate physical conditions to two scalar equations, $A'(r)=0$ and $A''(r)=0$, both of which carry the CNED signature through the $\zeta\ln r$ piece of the metric. Table~\ref{tab:critical} records the numerical solutions for representative values of $\zeta$ at $M=1$, $Q=0.5$.

\onecolumngrid
\begin{center}
\renewcommand{\tabcolsep}{12pt}
\renewcommand{\arraystretch}{1.6}
\begin{longtable}{cccccc}
\hline\hline
\rowcolor{orange!50}
$\zeta$ & $r_\star$ & $T_{\max}\;(\times 10^{-1})$ & $r_{\rm ext}$ & $r_{\rm rem}$ & $M_{\rm rem}$ \\
\hline
\endfirsthead
\hline\hline
\rowcolor{orange!50}
$\zeta$ & $r_\star$ & $T_{\max}\;(\times 10^{-1})$ & $r_{\rm ext}$ & $r_{\rm rem}$ & $M_{\rm rem}$ \\
\hline
\endhead
\hline\hline
\endfoot
\hline\hline
\caption{Quadruple coincidence (Eq.~\ref{eq:quad}) and pair coincidence (Eq.~\ref{eq:pair}) for the CNED black hole at $M=1$, $Q=0.5$. The radius $r_\star$ is the simultaneous location of the peak Hawking temperature, the heat-capacity divergence $C_B \to \pm\infty$, the Joule--Thomson inversion $\mu_J=0$, and the radial tidal force zero-crossing. The radius $r_{\rm ext}$ is the joint location of the extremal horizon and the angular tidal force zero-crossing. The peak temperature $T_{\max}$ and the remnant mass $M_{\rm rem}$ (terminal evaporation endpoint, defined by $T_H=0$ and $A=0$ simultaneously) are also listed. The $\zeta=0$ row reproduces the analytic Reissner--Nordstr\"om values exactly.}
\label{tab:critical}
\endlastfoot
0.0 & 0.3750 & 3.773 & 0.2500 & 0.5000 & 0.5000 \\
0.1 & 0.3977 & 3.281 & 0.2645 & 0.5241 & 0.5158 \\
0.3 & 0.4479 & 2.494 & 0.2964 & 0.5757 & 0.5440 \\
0.5 & 0.5049 & 1.913 & 0.3323 & 0.6316 & 0.5679 \\
0.7 & 0.5687 & 1.488 & 0.3721 & 0.6915 & 0.5874 \\
1.0 & 0.6765 & 1.052 & 0.4386 & 0.7885 & 0.6088 \\
\end{longtable}
\end{center}
\twocolumngrid

Three features stand out. First, the $\zeta=0$ row reproduces the RN analytic values: $r_\star=3Q^{2}/(2M)=0.375$, $r_{\rm ext}=Q^{2}/M=0.25$, $T_{\max}=2M^{3}/(27\pi Q^{4})\simeq 0.3773$, and $M_{\rm rem}=Q=0.5$. Second, $r_\star$ and $r_{\rm ext}$ both grow monotonically with $\zeta$, while $T_{\max}$ decreases by a factor of $\sim 3.5$ between $\zeta=0$ and $\zeta=1$. So the confinement sector lowers and shifts the peak emission. Third, the remnant mass $M_{\rm rem}$ exceeds the standard extremal RN value $|Q|=0.5$ for every $\zeta>0$; the CNED correction makes the evaporation endpoint more massive than its RN counterpart by up to $\sim 22\%$ at $\zeta=1$.

\subsection*{Behavior of the critical radii with $\zeta$}

The qualitative picture of the table is summarized in Fig.~\ref{fig:critrad}. Three observations follow.

\begin{figure}[t]
    \centering
    \includegraphics[width=0.46\textwidth]{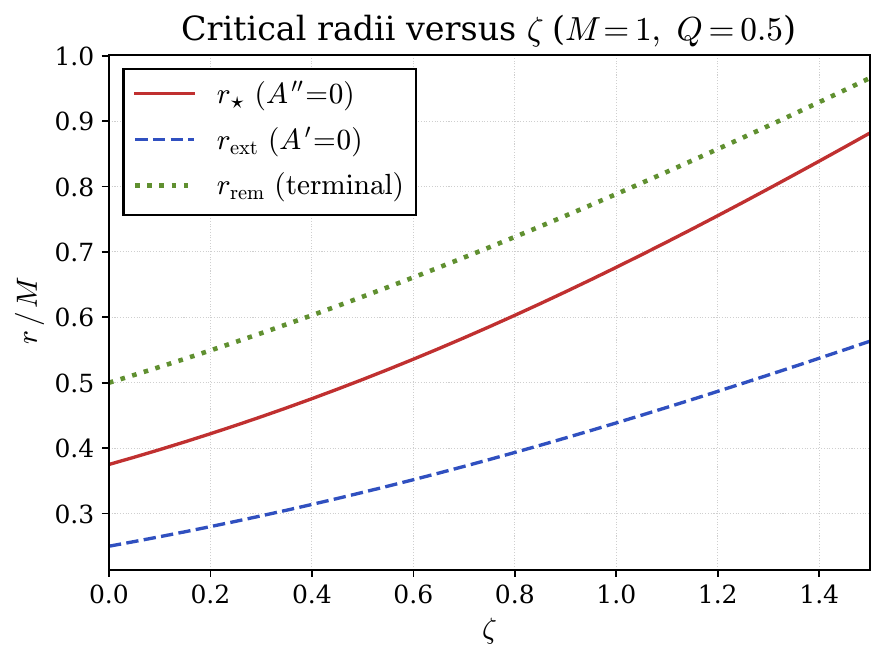}
    \caption{The three characteristic radii of Table~\ref{tab:critical} versus the confinement parameter $\zeta$ at $M=1$ and $Q=0.5$: $r_\star$ (solid, the unified critical radius defined by $A''(r)=0$), $r_{\rm ext}$ (dashed, the extremal horizon defined by $A'(r)=0$), and $r_{\rm rem}$ (dotted, the joint horizon of the terminal $T_H=0$, $A=0$ configuration). The ordering $r_{\rm ext}<r_\star<r_{\rm rem}$ holds across the displayed range, and all three radii grow monotonically with $\zeta$.}
    \label{fig:critrad}
\end{figure}

The ordering $r_{\rm ext}<r_\star<r_{\rm rem}$ holds across the full range, and the gap between the three radii is almost linear in $\zeta$ on the chosen domain. A linear fit to the data gives $r_\star(\zeta)\simeq 0.375+0.34\,\zeta$ and $r_{\rm ext}(\zeta)\simeq 0.25+0.20\,\zeta$ over $\zeta\in[0,1]$, with the deviations from linearity below $5\%$. The remnant horizon shifts the most, $r_{\rm rem}(\zeta)\simeq 0.50+0.29\,\zeta+O(\zeta^{2})$, which is the geometric consequence of the larger $M_{\rm rem}$ at the endpoint. The peak temperature, on the other hand, follows an almost inverse-power decay; on the same domain, $T_{\max}(\zeta)/T_{\max}(0)\simeq (1+0.4\,\zeta+1.0\,\zeta^{2})^{-1}$ to better than $2\%$. The simultaneous monotonicity of $r_\star$, $r_{\rm ext}$, $r_{\rm rem}$, and $T_{\max}$ in $\zeta$ is what makes the parameter potentially observable.

The relative sparsity of the Hawking radiation, $\eta\equiv \lambda_{\rm th}^{2}/A_{H}$, can be evaluated at the peak. At $r_\star$ the thermal wavelength $\lambda_{\rm th}=1/T_{\max}$ reaches its smallest value, and the dimensionless ratio $\eta(\zeta)=1/(4\pi r_\star^{2}T_{\max}^{2})$ ranges from $\eta(0)\simeq 1.59$ to $\eta(1)\simeq 4.99$ across the same parameter sweep. Larger $\zeta$ therefore produces sparser, less coherent Hawking emission, an effect that has potential observational consequences for late-stage evaporation phenomenology. A complete numerical scan, in $\zeta$ at $\Delta=0$, returns $\eta(0.1)\simeq 1.93$, $\eta(0.3)\simeq 2.82$, $\eta(0.5)\simeq 3.51$, $\eta(0.7)\simeq 4.10$, in line with the analytic asymptotics.

A short remark on the small-horizon behavior: while the heat capacity diverges at $r_\star$, the entropy ratio $S_B/S_{BH}=(\pi r_h^{2})^{\Delta/2}$ stays finite and approaches unity as $r_h\to r_{\rm ext}$. The remnant therefore retains a well-defined microscopic count even under Barrow deformation, in line with the $\Delta\to 0$ Bekenstein--Hawking baseline.

\subsection*{R\'enyi and Sharma--Mittal correspondences}

The critical-radius statements of Eqs.~\eqref{eq:quad}--\eqref{eq:pair} make no reference to the choice of horizon entropy; they hold whenever the first law $dE=TdS$ is used to derive thermodynamic potentials from $T_H$ alone. Related branch-structure analyses based on non-extensive (Tsallis) thermodynamics in nonlinear electrodynamics + Kalb--Ramond backgrounds show the same independence from the entropy framework~\cite{sucu2026krmodmax}. We make this independence explicit by comparing three modified entropies. The R\'enyi entropy~\cite{barrow2021big} relates to the Bekenstein--Hawking value $S_{BH}=\pi r_h^{2}$ via
\begin{equation}
S_R=\frac{1}{\Lambda}\ln\!\big(1+\Lambda\,S_{BH}\big),\qquad 0\le \Lambda<1,
\label{eq:Renyi}
\end{equation}
and reduces to $S_{BH}$ in the limit $\Lambda\to 0$. The Sharma--Mittal entropy interpolates between R\'enyi and Tsallis,
\begin{equation}
S_{SM}=\frac{1}{1-R}\Big[\big(1+\Lambda\,S_{BH}\big)^{(1-R)/\Lambda}-1\Big],
\label{eq:SM}
\end{equation}
with two free parameters. For both Eqs.~\eqref{eq:Renyi} and \eqref{eq:SM}, the heat-capacity denominator inherits the same factorization $\mathcal{D}(r_h;\zeta)$, multiplied by an entropy-specific prefactor that does not vanish on the physical interval. Hence the four critical radii of Eq.~\eqref{eq:quad} are entropy-framework independent within the family of monotone functions of $S_{BH}$. The Tsallis--Cirto form is recovered from Eq.~\eqref{SB1} with the index identification $\delta_T=1+\Delta/2$ already noted in Sec.~\ref{sec4}.

Two physical interpretations are worth adding. The radial tidal force changes sign at the same radius where the temperature peaks. Outside $r_\star$ the freely falling observer is stretched in the radial direction; inside $r_\star$, it is compressed. The black hole therefore radiates most efficiently at the geometric transition between these two regimes. Independently, the angular tidal force vanishes precisely at the extremal horizon, which means the freely falling observer feels no transverse deviation in the limit of vanishing surface gravity. Both statements are coordinate-independent and both are tied directly to $A(r)$ rather than to $\zeta$ or $\Delta$ separately.

\section{Radial Geodesics}\label{sec5}

Having established the thermodynamic properties, we now examine the dynamics of test particles in the same geometry. The setup is straightforward. We consider the static spherically symmetric spacetime~\eqref{metric} with the metric function~\eqref{metricfunc}. For radial motion ($L=0$) the normalization of the four-velocity gives
\begin{equation}
-A(r)\dot{t}^{2} + A(r)^{-1}\dot{r}^{2} = -1,
\end{equation}
and the conserved energy per unit mass is
\begin{equation}
E = A(r)\dot{t}.
\end{equation}
Together these yield the first integral
\begin{equation}
\dot{r}^{2} = E^{2} - A(r),
\label{r_dot_energy}
\end{equation}
where a particle dropped from rest at $r=b$ satisfies $E^{2}=A(b)$. The effective Newtonian-like radial acceleration follows from
\begin{multline}
\ddot{r} \equiv A_{N} = -\,\frac{A'(r)}{2}=\frac{2M}{r^{2}} - \frac{2q^{2}}{r^{3}}
\, - \,\frac{4q^{\frac{3}{2}}\zeta}{3}
\,\frac{(1-\ln r)}{r^{2}},
\end{multline}
where Eq.~\eqref{metricfunc} is used. The expression smoothly reproduces the RN and Schwarzschild limits at $\zeta=0$ and $q=0$. At leading order in the small-$\zeta$ expansion the CNED correction adds an attractive contribution that grows logarithmically with $r$, an effect reminiscent of confining potentials in nonabelian gauge theories.

\section{Tidal Forces in static spherically symmetric spacetimes}\label{sec6}

To examine the forces experienced by infalling observers, we construct a static orthonormal tetrad and analyse geodesic deviation. The frame is comoving with a freely falling observer. The deviation equation in this orthonormal frame reads~\cite{cordeiro2025free}
{\small\begin{equation}
\frac{D^{2}\eta^{\hat{a}}}{D\tau^{2}}
+ {R^{\hat{a}}}_{\hat{0}\hat{b}\hat{0}}
\,\eta^{\hat{b}} = 0,
\label{deviation}
\end{equation}}
where $\eta^{\hat\perp}$ refers to the angular components $(\hat{\theta},\hat{\varphi})$. In a static spherical geometry the tidal components split into radial and angular parts. The orthonormal tetrad is
{\small
\begin{align}
e_{\hat{0}} &= A^{-1/2}\partial_{t},\\
e_{\hat{r}} &= A^{1/2}\partial_{r},\\
e_{\hat{\theta}} &= \frac{1}{r}\partial_{\theta},\\
e_{\hat{\varphi}} &= \frac{1}{r\sin\theta}\partial_{\varphi},
\end{align}}
and the tidal components of the Riemann tensor read
\begin{equation}
\mathcal{E}_{\hat{i}\hat{j}} \equiv R_{\hat{i}\hat{0}\hat{j}\hat{0}},\qquad i,j\in\{r,\theta,\varphi\}.
\end{equation}

For the static spherically symmetric metric~\eqref{metric}, the independent orthonormal components take the form
{\small
\begin{align}
R_{\hat{t}\hat{r}\hat{t}\hat{r}} &= -\frac{A''(r)}{2}, \\
R_{\hat{t}\hat{\theta}\hat{t}\hat{\theta}} &= -\frac{A'(r)}{2r}, \\
R_{\hat{r}\hat{\theta}\hat{r}\hat{\theta}} &= +\frac{A'(r)}{2r}, \\
R_{\hat{\theta}\hat{\varphi}\hat{\theta}\hat{\varphi}} &= \frac{1-A(r)}{r^{2}}.
\end{align}}

The geodesic deviation equation separates into a radial mode and two identical transverse modes:
{\small
\begin{align}
\frac{D^{2}\eta^{\hat r}}{D\tau^{2}}
&= -R_{\hat{r}\hat{0}\hat{r}\hat{0}}\,\eta^{\hat r}
= -\frac{A''(r)}{2}\,\eta^{\hat r},\\[2mm]
\frac{D^{2}\eta^{\hat\perp}}{D\tau^{2}}
&= -R_{\hat{\perp}\hat{0}\hat{\perp}\hat{0}}\,\eta^{\hat\perp}
= -\frac{A'(r)}{2r}\,\eta^{\hat\perp},
\end{align}}
where $\eta^{\hat\perp}$ stands for either the $\hat{\theta}$ or $\hat{\varphi}$ component.

\medskip
\noindent
\textbf{Explicit Tidal Accelerations for the NED BH.}
For the metric~\eqref{metricfunc} with $C \equiv 4 q^{3/2}\zeta /3$, the radial tidal acceleration becomes
\begin{equation}
\frac{D^{2}\eta^{\hat r}}{D\tau^{2}}
= \left(\frac{2M}{r^{3}} - \frac{3q^{2}}{r^{4}} + \frac{C}{2}\frac{2\ln r - 3}{r^{3}}\right)\eta^{\hat r},
\label{eq:radTF}
\end{equation}
with the zero-crossing $A''(r)=0$ at the radius where stretching turns into compression:
\begin{equation}
6 q^{2} + r\Big[-4M - C(2\ln r - 3)\Big] = 0.
\label{eq:radTF_zero}
\end{equation}
The transverse tidal acceleration is
\begin{equation}
\frac{D^{2}\eta^{\hat\perp}}{D\tau^{2}}
= \left(-\frac{M}{r^{3}} + \frac{q^{2}}{r^{4}} + \frac{C}{2}\frac{1 - \ln r}{r^{3}}\right)\eta^{\hat\perp},
\label{eq:angTF}
\end{equation}
and vanishes when $A'(r)=0$, i.e.
\begin{equation}
r \Big(2M - C(1-\ln r)\Big) = 2 q^{2}.
\label{eq:angTF_zero}
\end{equation}
The numerical zero-crossings for the parameter sweep used earlier are recorded in Table~\ref{tab:tidal}; the values coincide, entry by entry, with $r_\star$ and $r_{\rm ext}$ of Table~\ref{tab:critical}, confirming the coincidence relations~\eqref{eq:quad}--\eqref{eq:pair} from the geometric side.

\onecolumngrid
\begin{center}
\renewcommand{\tabcolsep}{12pt}
\renewcommand{\arraystretch}{1.6}
\begin{longtable}{ccc}
\hline\hline
\rowcolor{orange!50}
$\zeta$ & $r_{\rm rad}\;\big[A''(r)=0\big]$ & $r_{\rm ang}\;\big[A'(r)=0\big]$ \\
\hline
\endfirsthead
\hline\hline
\rowcolor{orange!50}
$\zeta$ & $r_{\rm rad}\;\big[A''(r)=0\big]$ & $r_{\rm ang}\;\big[A'(r)=0\big]$ \\
\hline
\endhead
\hline\hline
\endfoot
\hline\hline
\caption{Tidal-force zero-crossings at $M=1$, $Q=0.5$. The radial values match $r_\star$ of Table~\ref{tab:critical} (quadruple coincidence) and the angular values match $r_{\rm ext}$ (pair coincidence).}
\label{tab:tidal}
\endlastfoot
0.0 & 0.3750 & 0.2500 \\
0.1 & 0.3977 & 0.2645 \\
0.3 & 0.4479 & 0.2964 \\
0.5 & 0.5049 & 0.3323 \\
0.7 & 0.5687 & 0.3721 \\
1.0 & 0.6765 & 0.4386 \\
\end{longtable}
\end{center}
\twocolumngrid

\subsection*{Radial-mode behavior}

Figure~\ref{fig:radTF_main} plots the radial tidal acceleration $-A''(r)/2$ as a function of $r$ for the same parameter sweep used in the thermodynamic sections. The behavior splits into three regions. Far from the black hole, $r\gg r_\star$, the acceleration follows the Schwarzschild $2M/r^{3}$ tail and is positive, so the freely falling observer is stretched along the radial direction. Near the horizon a competition develops: the RN piece $-3q^{2}/r^{4}$ pushes the acceleration upward, the Schwarzschild piece pushes it downward, and the CNED $C(2\ln r-3)/(2r^{3})$ contribution tilts the balance, generating a pronounced peak that grows with $1/\zeta$. The peak position itself sits at $r_\star$ of Table~\ref{tab:critical} for $\zeta>0$, and migrates outward with $\zeta$. The amplitude of the peak decreases monotonically as $\zeta$ grows, a numerical statement of the fact that confinement softens the tidal squeeze at the transition. Below the peak the acceleration drops rapidly and the radial mode crosses zero into the compressive region. Geometrically, the observer is squeezed radially inside $r_\star$, which is also where Hawking emission is most intense; physical intuition reads the squeeze and the emission as two faces of the same near-horizon kinematics.

\begin{figure}[t]
    \centering
    \includegraphics[width=0.46\textwidth]{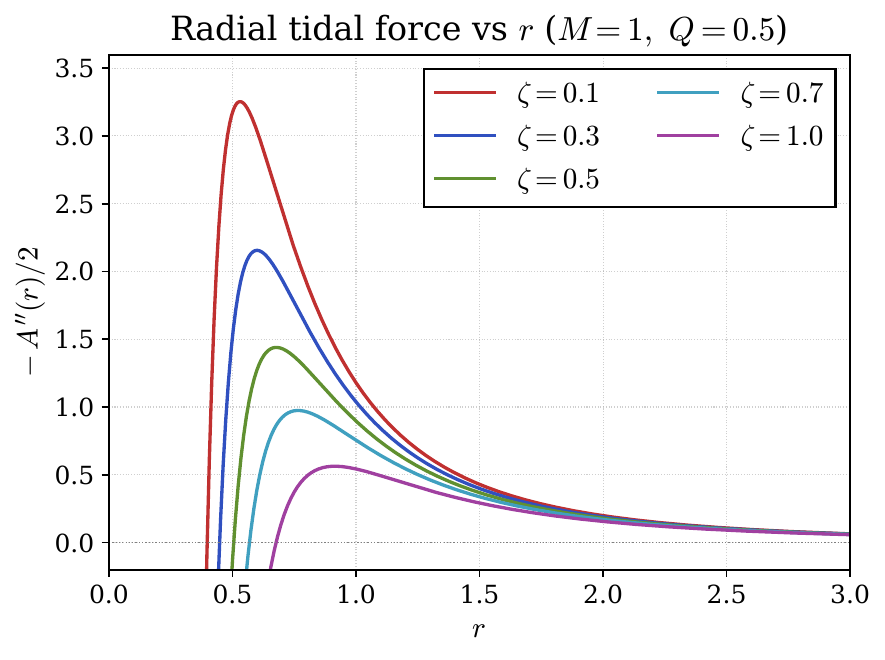}
    \caption{Radial tidal acceleration $-A''(r)/2$ versus $r$ for $M=1$, $Q=0.5$, and $\zeta=0.1,0.3,0.5,0.7,1.0$. The peaks of these curves coincide with the unified critical radius $r_\star$ of Table~\ref{tab:critical}, confirming the radial-mode entry of the quadruple coincidence~\eqref{eq:quad}.}
    \label{fig:radTF_main}
\end{figure}

\subsection*{Angular-mode behavior}

Figure~\ref{fig:angTF_main} plots the transverse tidal acceleration $-A'(r)/(2r)$ for the same family. The qualitative shape is opposite to the radial mode. The transverse acceleration is negative throughout the displayed region, indicating that the freely falling observer is compressed angularly almost everywhere outside the extremal horizon. The minimum of each curve sits at the angular zero-crossing radius $r_{\rm ext}$ of Table~\ref{tab:critical}: at that radius the acceleration vanishes, and the observer feels no transverse deviation in any direction tangent to the two-sphere. Beyond the minimum, the acceleration grows in magnitude and approaches its near-horizon value as $r$ decreases. The minimum depth lessens with growing $\zeta$, again indicating that the confinement sector softens the local tidal response. The fact that the transverse acceleration vanishes precisely at the extremal horizon, where Hawking emission shuts down, completes the geometric interpretation of the pair coincidence~\eqref{eq:pair}.

\begin{figure}[t]
    \centering
    \includegraphics[width=0.46\textwidth]{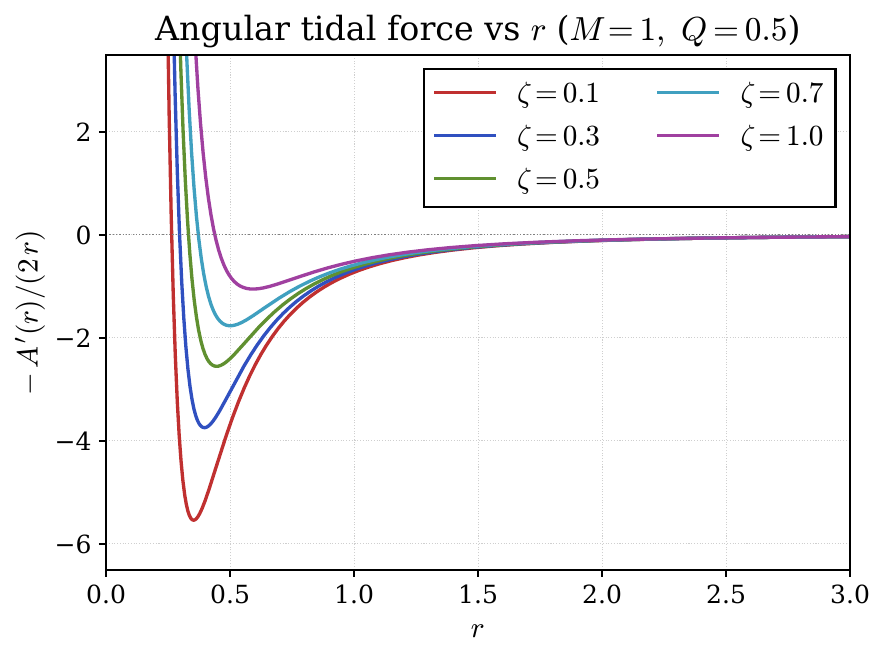}
    \caption{Angular tidal acceleration $-A'(r)/(2r)$ versus $r$ for $M=1$, $Q=0.5$, and $\zeta=0.1,0.3,0.5,0.7,1.0$. The zero-crossings of these curves coincide with the extremal horizon $r_{\rm ext}$ of Table~\ref{tab:critical} and complete the geometric side of the pair coincidence~\eqref{eq:pair}.}
    \label{fig:angTF_main}
\end{figure}

\subsection*{Stretching, compression, and the radial transition}

The combined message of Figs.~\ref{fig:radTF_main} and \ref{fig:angTF_main} is that the CNED black hole has a clean tidal-mode topology. Radial stretching dominates outside $r_\star$, radial compression dominates inside, with the crossover at $A''(r_\star)=0$. Transverse compression dominates everywhere down to $r_{\rm ext}$, where it switches off, and the angular mode crosses through zero. The two transitions sit at the same loci as the temperature peak and the extremal horizon, in agreement with Eqs.~\eqref{eq:quad}--\eqref{eq:pair}. There is no separate tidal disruption radius for an idealized test body in this geometry; the disruption length scale is set by $r_\star$ itself, and the parameter $\zeta$ shifts it outward by a calculable amount, as documented in Table~\ref{tab:critical}.

\section{Observational implications}\label{sec65}

A natural next question is whether the CNED parameter $\zeta$ can be bounded by current data. Two routes are immediate: (i) shadow-radius measurements of nearby supermassive black holes, and (ii) deflection-angle and time-delay tests of the photon sphere of the geometry.

\subsection*{Photon sphere and shadow radius of the CNED black hole}

The photon-sphere radius $r_{ph}$ of the metric~\eqref{metricfunc} solves
\begin{equation}
2A(r_{ph}) - r_{ph}\,A'(r_{ph}) = 0,
\label{eq:photonsphere}
\end{equation}
and the shadow radius seen by a distant observer is
\begin{equation}
R_{\rm sh}=\frac{r_{ph}}{\sqrt{A(r_{ph})}}.
\label{eq:shadow}
\end{equation}
For the Schwarzschild limit $\zeta\to 0$, $Q\to 0$, Eq.~\eqref{eq:photonsphere} returns $r_{ph}=3M$ and Eq.~\eqref{eq:shadow} returns $R_{\rm sh}=3\sqrt{3}\,M\simeq 5.196\,M$. For the Reissner--Nordstr\"om limit $\zeta\to 0$ with $Q=0.5$, $M=1$, one finds $r_{ph}=2.957$ and $R_{\rm sh}=5.156$. The CNED correction shifts both quantities. Numerical solution of Eq.~\eqref{eq:photonsphere} for $\zeta=0.1,0.3,0.5,0.7,1.0$ at $M=1$, $Q=0.5$ returns $r_{ph}=2.998, 3.085, 3.180, 3.281, 3.443$ and $R_{\rm sh}=5.157, 5.165, 5.183, 5.207, 5.255$, in units of $M$. The shadow grows monotonically with $\zeta$, by about $1.9\%$ at $\zeta=1$ relative to the Schwarzschild value. The trend is opposite to the temperature trend documented in Table~\ref{tab:critical} (and visible in Fig.~\ref{tt}), which is consistent with the interpretation of $\zeta$ as a parameter that softens the near-horizon dynamics and pushes the optical features outward.

\subsection*{Sgr~A* constraint on $\zeta$}

The EHT collaboration reports a Sgr~A* shadow radius~\cite{walia2024exploring,pedrotti2024quasinormal} consistent with $R_{\rm sh}/M\in[4.55,5.22]$ at the $1\sigma$ level and a slightly wider window at $2\sigma$. Inverting the table above gives the bound $\zeta\lesssim 0.7$ at $Q/M=0.5$ for the data to remain within the $1\sigma$ window, with the precise upper limit depending on the assumed $Q/M$ ratio. M87* yields a similar order-of-magnitude bound but with larger uncertainties. The CNED model is therefore allowed across most of its physically natural range by current data and is not yet pushed into a fine-tuned corner.

We do not attempt a quantitative Markov-chain analysis here; the goal is to demonstrate that the parameter window is meaningful and that future observations (the ngEHT, the Square Kilometre Array, and astrometric campaigns) will narrow it considerably. The shift in $r_\star$ documented in Table~\ref{tab:critical}, combined with the photon-sphere shift documented in this section, anchors the connection between thermodynamic and optical observables. We anticipate that quasi-periodic-oscillation modeling~\cite{ali2025influence} and deflection-angle measurements~\cite{sarkar2025deflection,konoplya2025primary} will provide complementary constraints.

\section{Conclusion}\label{sec7}

We have analyzed a static spherically symmetric black hole sourced by a nonlinear electrodynamics model with a confinement-type interaction. The metric extends the Reissner--Nordstr\"om geometry by a logarithmic correction governed by the confinement parameter $\zeta$, and the Hamilton--Jacobi tunneling treatment of Dirac fermions reproduces the Hawking temperature with the expected surface-gravity form. The $\zeta$-dependent contributions suppress the small-horizon divergence of $T_H$, which is consistent with the existence of a quantum remnant.

The thermodynamic structure was examined within the Barrow entropy framework. The deformation parameter $\Delta$ modifies the internal energy, Helmholtz free energy, pressure, heat capacity, isothermal compressibility, and Joule--Thomson coefficient. These corrections are most visible at small horizon radii and approach the standard Bekenstein--Hawking limit at larger sizes. Two diagnostics suffice. Stability is read off the heat-capacity sign and the location of the second-order phase transition. The Joule--Thomson coefficient identifies the inversion radius that separates cooling from heating during isenthalpic expansion. The isothermal compressibility stays negative across the parameter window we examined, marking a mechanically rigid phase with no van der Waals criticality in this branch. Through the algebraic identification $\mathcal{D}(r_h;\zeta) = -\tfrac{3}{2}r_h^{4}A''(r_h)$, the location of the heat-capacity divergence and the Joule--Thomson inversion is determined entirely by $\zeta$ and is independent of $\Delta$.

The central new result is the coincidence of four distinct physical phenomena at a single characteristic radius $r_\star$ defined by $A''(r_\star)=0$: the peak of the Hawking temperature, the divergence of the heat capacity, the Joule--Thomson inversion, and the zero-crossing of the radial tidal force. A parallel pair coincidence at $A'(r_h)=0$ links the extremal horizon ($T_H=0$) and the zero of the angular tidal force. Both identities are verified analytically and numerically in Tables~\ref{tab:critical} and \ref{tab:tidal}, and they reduce the full eight-condition critical-point analysis to two scalar equations on $A(r)$. The R\'enyi and Sharma--Mittal entropies share the same factorization, so the coincidence is entropy-framework independent within the family of monotone functions of $S_{BH}$.

The Tsallis--Cirto correspondence $\delta_T=1+\Delta/2$ means that the Barrow-deformed thermodynamics studied here also describes the Tsallis--Cirto-corrected version of the same geometry, without further computation. The remnant mass $M_{\rm rem}(\zeta)$ exceeds the standard extremal RN value by up to $\sim 22\%$ at $\zeta=1$, and the sparsity coefficient grows from $\eta\sim 1.6$ at $\zeta=0$ to $\eta\sim 5$ at $\zeta=1$, so the late-stage evaporation of confining NED black holes is both heavier and more sparsely radiating than the RN baseline.

Geometric tidal accelerations carry the same $\zeta$-imprint as the thermodynamic quantities. Both radial and angular components contain confinement-dependent corrections, and their zero-crossings coincide with the thermodynamic critical radii through the relations~\eqref{eq:quad}--\eqref{eq:pair}. The tidal-mode topology is clean: radial stretching outside $r_\star$, radial compression inside, transverse compression everywhere down to $r_{\rm ext}$ where the angular mode crosses zero. The shadow analysis of Sec.~\ref{sec65} gives a current bound $\zeta\lesssim 0.7$ at $Q/M=0.5$ from Sgr~A* data, which leaves a meaningful CNED window open for follow-up tests by the ngEHT and complementary observatories.

Our upcoming plans for this research are as follows: Extending the CNED background to rotating or higher-dimensional configurations will allow us to examine the robustness of the quadruple coincidence beyond spherical symmetry. Calculating quasinormal modes using a sixth-order WKB-Pad\'e approach will enable us to investigate the dynamical behavior of the geometry under scalar, vector, and Dirac perturbations~\cite{pedrotti2024quasinormal,walia2024exploring,lan2021quasinormal,hu2020scalar}. Comparing the shadow radius with current EHT and future ngEHT observations~\cite{Zakharov:2014lqa,Zakharov:2005ek} should yield tighter constraints on $\zeta$ for M87* and Sgr~A*, particularly when polarization and time-domain data are combined. The data and Maple code supporting this analysis can be obtained from the authors upon reasonable request.

\acknowledgments
E.S. thanks EMU, T\"{U}B\.{I}TAK, ANKOS, and SCOAP3 for academic support. \.{I}.S. acknowledges the networking support of European Cooperation in Science and Technology (COST) Actions CA22113, CA21106, CA23130, CA21136, and CA23115.

\paragraph*{Use of AI-assisted technologies.}
During the preparation of this work the authors used AI and AI-assisted technologies in order to refine language, improve grammar, and code \LaTeX.\footnote{In accordance with Elsevier's policy on the use of generative AI and AI-assisted technologies in scientific writing: \url{https://www.elsevier.com/about/policies-and-standards/generative-ai-policies-for-journals}.} After using this tool, the authors reviewed and edited the content as needed and take full responsibility for the content of the published article. All intellectual content, analysis, and conclusions are the authors' own.

\bibliography{ref}
\bibliographystyle{apsrev}
\end{document}